\documentclass{aastex}

\begin{document}

\title{Origin of coherent magnetic fields in high redshift objects}
\author{A. Kandus\altaffilmark{1}, R. Opher\altaffilmark{1}, and S. R. M.
Barros\altaffilmark{2}}

\altaffiltext{1}{IAG, Universidade de S\~ao Paulo, Rua do Mat\~ao 1226,\\
Cidade Universit\'aria, CEP 05508-900 S\~ao Paulo, SP, Brazil\\
E-mail: kandus@astro.iag.usp.br, opher@astro.iag.usp.br} 
\altaffiltext{2}{IME, Universidade de S\~ao Paulo, Rua do Mat\~ao 1010,\\
Cidade Universit\'aria, CEP 05508-900, S\~ao Paulo, SP, Brazil\\ 
E-mail: saulo@ime.usp.br} 

\shorttitle{Magnetic fields in high z objects} 

\shortauthors{Kandus, Opher, and Barros}

\begin{abstract}
Large scale strong magnetic fields in galaxies are generally thought to have been 
generated by a mean field dynamo. In order to have generated the fields observed,
the dynamo would have had to have operated for a sufficiently long period of time. 
However, magnetic fields of similar intensities and scales to the one in our galaxy, are
observed in high redshift galaxies, where a mean field dynamo would not have had
time to produce the observed fields. Instead of a mean field dynamo, we study the 
emergence of strong large scale magnetic 
fields in the first objects formed in the universe due to the action of a turbulent,
helical stochastic dynamo, for redshifts $5 \leq z \leq 10$. Ambipolar 
drift plays an important role in this process due to the low level of 
ionization of the gas, allowing a large scale stochastic dynamo to operate. 
We take into account the 
uncertainties in the physics of high redshift objects by examining a range of 
values for the parameters that characterize the turbulent plasma. By
numerically integrating the nonlinear evolution equations for magnetic field
correlations, we show that for reasonable values of the parameters
in the time interval considered, fields can grow to high intensities 
($\sim 10^{-6}$ G), with large coherence lengths ($\sim 2-6$ kpc), 
essentially independent of the initial values of the magnetic field.
\end{abstract}


\section{INTRODUCTION}

Magnetic fields have been observed in all known structures of our universe, from
the Earth to superclusters of galaxies, spanning a wide range of
intensities from $\sim\mu\,\rm{G}$ in galaxies and galaxy clusters to $\sim
10^{\,12}\,\rm{G}$ in neutron stars. The origin of these fields in large structures,
such as galaxies and clusters of galaxies, remains an unsolved
problem. Many physical processes have been proposed to explain the
origin and evolution of these fields \citep{prim-b,widrow}.
The processes suggested can be divided into two main classes: 1) cosmological mechanisms; 
and 2) local astrophysical processes. Until now, none of them 
has provided a satisfactory explanation for the generation of the magnetic fields.
\par
A mean field dynamo is commonly invoked to explain the fields observed in our 
galaxy and in small redshift galaxies (e.g., \citealp{zeldovich83,moffat78}). 
In order to have attained the observed intensities, the dynamo would have 
had to have operated for a time on the order of the age of the universe.
However, the presence of equally intense and coherent fields in high
redshift galaxies \citep{carrilli02}, where the mean field dynamo would not
have had enough time to amplify the field to the observed values, casts doubt on the
mean field dynamo paradigm as the preferred generation mechanism. Fields of
similar intensity and coherence to those in the Milky Way, have also been detected in high
redshift damped $\rm{Lyman}-\alpha $ systems \citep{oren}.
\par
In this paper, we are concerned with the origin of strong, large coherent magnetic fields
in high redshift structures. We relate the origin of magnetic fields to
the physical conditions in the early universe: 1) a dense low ionized
plasma; and 2) appreciable turbulence due to the observed high star formation rate
(Lanzetta et al. \citealp{lanzetta02}).
\par
The formation of the first stars and quasars marked the beginning of the
transformation of the universe from a smooth initial state to its present 
lumpy state. In the bottom-up hierarchy of cold dark matter (CDM)
cosmologies, the first gaseous clouds collapsed at redshifts $z > 10$ and, 
subsequently, fragmented into stars due to molecular hydrogen cooling 
\citep{loeb01}. These collapsing objects then fragmented into many clumps, which
had typical masses of $\sim 10^{\,2}-10^{\,3}M_{\odot }$. Very massive stars have
lifetimes of $\sim 3\times 10^{\,6}\,\rm{years}$ and end their lives as
supernovae. 
\par 
Recently, Lanzetta et al. \citep{lanzetta02} showed that
the incidence of the highest intensity star formation regions increases
monotonically with redshift. Their observations indicate that star formation in the early
universe occurred at a much higher rate than was previously believed. Therefore,
the rate of occurrence of supernovae would have also been much higher in the past
than at the present. Supernovae shocks disturb the plasma in which they are
immersed, producing turbulent motions of the gas. If the supernovae rate was much
higher in the past than at present, the plasma of the first
formed objects must have been much more turbulent than that of presently observed, low
redshift, star forming galactic molecular clouds. 
\par 
Turbulence generates stochastic
magnetic fields (magnetic noise) at a faster rate than it does mean
fields \citep{kul-and-92}. If the turbulence is strongly non-helical, 
the fields induced are confined to small scales (Kazantzev \citealp{kazantzev68}). 
However, if it is helical, induction of large scale magnetic correlations by 
the $\alpha$ effect occurs
(Vainshtein \& Kichatinov \citealp{vainshtein86}). Astrophysical turbulence is mainly of a
helical nature. Hence, we can expect that large scale correlations were
induced by the high redshift, turbulent plasma.
\par
In this study, we explore the hypothesis that the magnetic fields observed in
high redshift galaxies were created by small scale, stochastic, turbulent
helical dynamos, rather than by mean field dynamos. We use a simple model of a
gas cloud that is assumed to have collapsed at a high redshift $z > 10.$  At 
$z \sim 10,$ the cloud would have had a low magnetization level and a high 
level of turbulence. Thus, it would have been similar to the turbulent, low 
ionization, star forming molecular
clouds observed in our galaxy, albeit with a much smaller initial magnetic
field and a much higher turbulence level due to the higher star formation
rate in the early universe.
\par
It is well known that shock waves produced by supernova explosions
accelerate cosmic rays to energies $\geq\,\rm{GeV}$ (e.g., \citealp{cr-acc}). V\"{o}lk et
al. \citep{volk89} showed that in all galaxies, the supernova rate is a
direct measure of the cosmic ray intensity. We can, therefore, infer that
cosmic rays were already present in considerable intensities in high
redshift galaxies. We take into account phenomenologically, the effect on turbulence of
cosmic rays, supernova shocks, and powerful stellar winds from massive
stars on turbulence by varying the turbulent parameters over a broad range,
in order to take into account the uncertainies
in our knowledge of high redshift structures.
\par
The linear evolution equations for the correlation function of magnetic fields 
for non-helical turbulence were
derived nearly forty years ago by Kazantzev \citep{kazantzev68}. 
For helical turbulence, the corresponding equations
were obtained twenty years later by Vainshtein \& Kichatinov \citep{vainshtein86}. 
These equations are linear in the magnetic correlations.
Recently, Subramanian \citep{subramanian99} and Brandemburg \& Subramanian \citep{bran-sub}
derived the non-linear evolution
equations for the magnetic correlations by taking into account the
back-reaction of the Lorentz force on the plasma charges in the form of
ambipolar drift. We solve the nonlinear helical
evolution equations numerically for various values of the parameters that
characterize the high redshift turbulent plasma.
\par
The paper is organized as follows. In section II, we give the  evolution
equations for the magnetic correlations. We describe the effects of the
main parameters on the integration in section III . Finally, in section IV, 
we summarize and discuss
our results.

\section{MAGNETIC FIELD EVOLUTION EQUATIONS}

In this section, we summarize the nonlinear evolution equations for the magnetic field
correlations  (Subramanian \citealp{subramanian99}, Brandemburg \& Subramanian 
\citealp{bran-sub}). The evolution equation for the magnetic
field is given by the induction equation, $\partial \,\mathbf{B}/\partial\, t=\mathbf{
\nabla \times} \left( \mathbf{v\times {B}}
-\eta \mathbf{\nabla
\times \mathbf{B}}\right),$ where $\mathbf{B}$ is the magnetic
 field, 
$\mathbf{v}$ the velocity of the fluid, and $\eta$ is the Ohmic resistivity. The
velocity $\mathbf{v}$ $(=\mathbf{v}_{\,T}+\mathbf{v}_{D})$ is the sum of an external
stochastic field $\mathbf{v}_{\,T}$ and an ambipolar drift component $\mathbf{v}_{D},$
which describes the non-linear back-reaction of the
Lorentz force. This back-reaction is due to the force that the ionized gas exerts on 
the neutral gas through collisions of the ions with
the neutral atoms. It is assumed that $\mathbf{v}_{\,T}$ is an isotropic,
homogeneous, Gaussian random field with a zero mean value and a delta correlation 
function in time (Markovian approximation). Its two point correlation
function is $\left\langle {v_{\,T}^{\,i}\left( \mathbf{x},t\right) v_{\,T}^{\,j}\left( 
\mathbf{y},s\right)} \right\rangle = T^{\,ij}(\mathbf{r})\delta(t-s),$ where
$T^{\,ij}\left( \mathbf{r}\right)
=T_{NN}\left( \mathbf{r}\right) \left[\, \delta ^{\,ij}- 
{r^{\,i}r^{\,j}}/{r^{\,2}}\,\right]
+T_{LL}\left( \mathbf{r}\right) \left( {r^{\,i}r^{\,j}}/{r^{\,2}}\right) +C\left( 
\mathbf{r}\right) \epsilon ^{\,ijf}r_{f}$ \citep{monin75}. The symbol $\left\langle 
{}\right\rangle $
denotes ensemble averaging over the stochastic velocities, 
$r=\left| \mathbf{x}-\mathbf{y}\right|,$ $r^{\,i}=x^{\,i}-y^{\,i},$ 
$T_{LL}\left( r\right) $ and $T_{NN}\left( r\right)$ 
are the longitudinal and transverse correlation functions of the velocity
field, respectively, and $C\left( r\right) $ is the helical term of the
velocity correlations. As the magnetic field grows, the Lorentz force acts
on the fluid. We assume that the fluid responds instantaneously  
and develops an extra drift velocity, proportional to
the instantaneous Lorentz force. We, thus, express the drift velocity as 
$\mathbf{v}_{D}=a\left[ \left( 
\mathbf{\nabla \times B}\right) \times \mathbf{B}\right],$ where 
$a=\tau /4\pi \rho _{\,i}$, $\tau $ the characteristic response time, and 
$\rho_{\,i}$ is the ion density.
\par
Consider a system whose size $S\gg L_c,$ where $L_c$ is the coherence scale
of the turbulence, for which the mean field averaged over any scale is
negligible. We take $\mathbf{B}$ to be a homogeneous, isotropic, Gaussian
random field with a negligible mean average value. Thus, we take the equal time, two
point correlation of the magnetic field as 
\begin{equation}
\left\langle B^{\,i}\left( \mathbf{x},t\right) B^{\,j}\left( \mathbf{y}
,t\right) \right\rangle =M^{\,ij}\left( r,t\right),
\end{equation}\label{c}
where 
\begin{equation}
M^{\,ij}=M_{N}\left[\delta ^{\,ij}-\left( \frac{r^{\,i}r^{\,j}}{r^{\,2}}\right) 
\right] +M_{L}\left( \frac{r^{\,i}r^{\,j}}{r^{\,2}}\right) +H\epsilon _{\,ijk}\,r^{\,k}
\label{d}
\end{equation}
(Subramanian \citealp{subramanian99}). The symbol $\left\langle {}\right\rangle $ 
denotes a double ensemble average over
both the stochastic velocity and $\mathbf{B}$ fields, 
$M_{L}\left( r,t\right) $ and $M_{N}\left( r,t\right) $ are the longitudinal
and transverse correlation functions, respectively, of the magnetic field, and 
$H\left(r,t\right) $ is the helical term of the correlations. Graphically, 
$M_L$ can be represented as $\rightarrow --\rightarrow$ and $M_N,$ as 
$\uparrow --\uparrow.$ Hence, positive values of $M_L$ and $M_N$ correspond to
parallel vectors and negative values, to anti-parallel vectors. Since $
\mathbf{\nabla \cdot{B}}=0,$ we have $M_{N}=\left( 1/2\,r\right) \partial
\left( r^{\,2}M_{L}\right) /\left( \partial r\right) $ \cite{monin75}. The
induction equation can be converted into evolution equations for $M_{L}$ and 
$H:$
\begin{eqnarray}
\frac{\partial M_{L}}{\partial \,t}\left( r,t\right) &=&\frac{2}{r^{\,4}}\frac{
\partial }{\partial
\,r}\left( r^{4}\kappa _{N}\left( r,t\right) \frac{
\partial M_{L}\left( r,t\right) }{\partial \,r}\right)  \nonumber \\
&+& G(r)M_{L}\left( r,t\right) +4\,\alpha _{N}H\left( r,t\right),  \label{e}
\end{eqnarray}
\begin{eqnarray}
\frac{\partial H}{\partial {\,t}}\left( r,t\right) &=& \frac{1}{r^{\,4}}\frac{
\partial }{\partial {\,r}}\left[ r^{\,4}\frac{\partial }{\partial {\,r}}\left[ \,2\,\kappa
_{N}\left( r,t\right) H\left( r,t\right) \right. \right. \nonumber\\ 
&-& \left. \left. \alpha _{N}\left( r,t\right) M_{L}\left( r,t\right)\right]\, \right] , 
\label{f}
\end{eqnarray}
where 
\begin{equation}
\kappa _{N}\left( r,t\right) =\eta +T_{LL}\left( 0\right) -T_{LL}\left(
r\right) +2\,a\,M_{L}\left( 0,t\right)  \label{g},
\end{equation}
\begin{equation}
\alpha _{N}\left( r,t\right) =2\,C\left( 0\right) -2\,C\left( r\right)
-4\,a\,H\left( 0,t\right)  \label{h},
\end{equation}
and
\begin{equation}
G\left( r\right) =-4\left\{ \frac{d}{d\,r}\left[ \frac{T_{NN}\left( r\right) }{
r}\right] +\frac{1}{r^{\,2}}\frac{d}{d\,r}\left[\, r\,T_{LL}\left( r\right) \right]
\right\}  \label{i}
\end{equation}
(Subramanian \citealp{subramanian99}). These equations form a closed set of 
nonlinear partial differential
equations for the evolution of $M_{L}$ and $H,$ describing the evolution of
magnetic correlations at small and large scales. The effective diffusion
coefficient $\kappa _{N}$ includes microscopic diffusion $(\eta ),$ a scale-dependent 
turbulent diffusion $\left[T_{LL}\left( 0\right) -T_{LL}\left( r\right)\right], $ 
and ambipolar drift $2aM_{L}\left( 0,t\right) ,$ which is
proportional to the energy density of the fluctuating fields. Similarly, $
\alpha _{N}$ is a scale-dependent $\alpha $ effect, proportional to 
$[\,2\,C\left( 0\right)
-2\,C\left( r\right )].$ The nonlinear decrement of the $\alpha $ effect due to ambipolar 
drift is  $4aH\left(0,t\right), $ proportional to the mean helicity of the magnetic
fluctuations. The $G\left( r\right) $ term in equation (\ref{e}) allows for rapid 
generation of small scale magnetic fluctuations due to velocity shear 
(\citealp{zeldovich83}; Kazantzerv \citealp{kazantzev68}). 
We are interested in the evolution of $M_L\left(r \right)$ since this function 
gives information about the coherence of the
induced large scale magnetic field. A
positive value of this function over a given
length indicates that the field is coherent in this region. Therefore this
length will be taken as the coherence scale of the induced field. Since $M_L$ is the
correlation function of the tensor product of parallel vectors, evaluated at two points 
separated by a distance $r,$ we can estimate the induced magnetic field intensity at
all points where $M_L>0$ as $B\sim M_L(r)/M_L^{\,1/2}(0).$

\section{TURBULENT STOCHASTIC DYNAMO ACTION \\IN HIGH REDSHIFT GALAXIES}

In order to study the evolution of the magnetic correlations due to the
turbulent plasma in the high redshift objects, we integrated equations (\ref{e}) and 
(\ref{f})
numerically for different values of 
the parameters. We employed second order conservative finite differencing in space,
with Neumann boundary conditions. In the time discretization, we used a
second order Crank-Nicolson type method, except for the treatment
of the non-linear terms. In these  terms, we employed the values of $M_L(0,t)$ and
$H(0,t)$ from the previous time-step, making the 
system of equations to be solved, linear in each time-step. The equations were
solved by a few iterations of a relaxation procedure. 
The implicit treatment in the time discretization
is important to avoid the severe stability constraints that would result from a 
fully explicit time discretization of the system. 
For the numerical results presented in this paper, we 
employed a spatial grid with 5000 equally spaced grid-points. With this
resolution, we were able to obtain convergence.
Doubling the resolution led to graphically indistinguishable numerical
results.

\subsection{Characterizing the High Redshift Plasmas}

We considered a cloud at $z\sim 10$ and followed the evolution of the
magnetic correlations until $z\sim 5$ $(\sim 10^{\,9}\,
\rm{ years)}.$ The value
taken for the cut-off scale of the turbulence, $l_c\sim 1\;\rm{AU},$ is similar to
that for present objects \citep{zeldovich83}. Assuming $L_c \gg l_c,$ we studied the
range of values $10\,\rm{pc} \la L_{c}\la 100\,\rm{pc}.$ We assumed that the height 
$h$ of 
the turbulent eddies of the high
redshift object is of the same order of magnitude as $L_c$. In order to
estimate the correlation velocity $V_c$ on the scale $L_c,$ we used
the expression $V_c^{\,2} \left( V_c/L_c\right) \sim \varepsilon,$
where $\varepsilon $ is the turbulent energy dissipated per unit
mass per unit time. This expression assumes that the energy is dissipated on the 
order of a single
rotation of the eddies of size $L_c$ at the angular frequency $\Omega \sim
V_c\,/L_c.$ We then have $V_c \sim \left( \varepsilon L_c\right)
^{1/{\,3}}.$ Supernova explosions are a major contributor to the galactic
turbulent energy. The energy associated with a supernova remnant in our galaxy is
about $3\times 10^{\,50}\,\rm{ erg},$ with about one third 
transformed into kinetic energy of the ambient gas.
Larger values for the supernova remnant energy and the mass of the gas
involved in the explosions, will produce higher turbulent velocities.
We assumed that at
redshifts 5-10, $\,f$ explosions occurred every 5 years and that the mass
of the gas involved was $10^{\,10} M_{\odot}$ \cite{zeldovich83}. As
noted above, the star formation and supernova rates were very high in
the past. The indicated star formation rate from observations increased by a 
factor of $\sim 50, $ in going from $z\sim 0$ to $z\sim 8$ (see e.g., fig. 4 
in Lanzetta et al.
\citealp{lanzetta02}). The expected values for $f$ are then $1 < f\la 10.$ 
A value of $f\sim 0.1$ corresponds to the present supernova rate in our galaxy.
We, thus, have $\varepsilon \simeq 0.3\times f\;\rm{cm^{\,2}\,s^{-3}}.$ For the
considered values of $L_c,$ the expected range of values
for $V_c$ is $9.59\,{\rm{km\,s^{-1}}} \la V_c \la 96.5\,\rm{km\,s^{-1}}.$ These 
values are 3 - 10 times larger than those in our galaxy \cite{zeldovich83}.
Assuming that the largest velocity corresponds to the largest
eddy, we have $\Omega \sim 10^{-13}\,\rm{ s^{-1}}.$ We estimated that the baryon
density is  $\rho_n\left( z\right) = \rho_n\left(
0\right)\left( 1+z\right)^{\,3} b,$ where $\rho_n\left( 0\right)$ is the
present baryon density and $b$ is a compression factor, which can be much
greater than $\sim 200$ (virial collapse). In our galaxy, the particle
density is $\sim 1\,\rm{ cm^{-3}}$ or $\rho_n \sim 10^{-24}\,\rm{g\,cm^{-3}}.$ 
The average baryon density in the universe today is $\sim 10^{-30}\,\rm{g\,cm^{-3}}.$
Thus, for our galaxy, the compression factor is $b\sim 10^{\,6}.$ We assumed that the
cloud that we are studying in the interval $5 \leq z\leq 10,$ collapsed 
virially at a high redshift, creating a large $b.$
Reasonable values for $b$ are, then, in the range $200 \leq b \leq 10^{\,7}.$
Taking $\rho_n\left( 0\right) \sim 0.05\,\rho_c\left( 0\right),$ where 
$\rho_c\left( 0\right) \simeq 0.9\times 10^{-29}\,\rm{g \,cm^{-3}}$ is the present
critical density (assuming a fiducial factor, $h\sim 0.7$, for the Hubble
constant), we obtain $4\times 10^{-26}\,\rm{g \,cm^{-3}} \la \rho_n$ $(z=10) \la 
2.3 \times 10^{-21}\,\rm{g\,cm^{-3}}$ for the baryon density in our high redshift
cloud. We estimated the ion mass density as $\rho_i \sim g\rho_n, $ with 
$0.001 \la g \la 1,$ which gives an ion density in the range $4\times
10^{-29}\rm{g\,cm^{-3}} \la \rho_i \la 2.3 \times 10^{-21}\, \rm{g\,cm^{-3}}.$
\par
At $z\sim 10,$ the cosmic microwave radiation temperature was 
$\left( 1 + z\right) T_0 \sim 30\,\rm{K}.$ For $5\la z\la 10,$ we considered 
plasma cloud temperatures in the interval $30\,\rm{K} \la T\la 10^{\,3}\,\rm{K}.$ Using
these values and estimating the thermal velocity of the ions as $v_n = (3k_B
T/m_p)^{1/2},$ we obtained $10^{\,4} \,{\rm{cm\,s^{-1}}}\la v_n\la 10^{\,5}\, 
\rm{cm\,s^{-1}}.$
Comparing these values with $V_c,$ we see that we are dealing with mildly supersonic
turbulence.
\par
Due to the the relatively low temperatures of the plasma, the ion-neutral
collision cross section is $\sigma_{in}\simeq 10^{-15}\,\rm{cm^{\,2}}$ 
\citep{NRL}. The ion-neutral collision frequency is $\nu_{in} = \sigma_{in}\,
n_n \,v_{th}$, giving $10^{-16} \,s^{-1} \la \nu_{in}\la 10^{-10} s^{-1}.$ The 
electrical resistivity can be estimated as $\eta = \left(
c^{\,2/4}\pi\right) \left(m_e\,\nu_{en}/e^{\,2}\,n_e\right),$ where $n_e$ is the electron
number density, $m_e$ the electron mass, and $\nu_{en} =
\langle\sigma_{en}\,v_e\,\rangle \,n_n$ is the electron-neutral collision frequency.
Taking $n_e = n_i$ (charge neutrality), $T_e \sim T_i,$ and
using $v_e \sim \left( 3k_BT_e/m_e\right)^{\,1/2},$ we obtain $\eta \sim
5\times 10^{\,3}\,\rm{cm^{\,2}\,s^{-1}},$ which is extremely small. The magnetic
Reynolds number is $R_m = L_cV_c\,/\eta \sim 10^{\,23} - 10^{\,24},$ which means
that at high redshifts, plasma turbulence was the main mechanism for
diffusion and dissipation. Thus the first term in equation (\ref{g}) can be
neglected. Since the ion-neutral collision was the dominant interaction in the
plasmas considered, we took the characteristic response time as $\tau \sim
\nu_{in}^{-1}.$ The coefficient $``a"$ in the non-linear terms in equations (\ref
{g}) and (\ref{h}) can then assume values in the interval $4.3 \times 10^{\,30}\;
{\rm{g^{-1}\,cm^3\,s}} \la a \la 2.5 \times 10^{\,
44}\;\rm{g^{-1}\,cm^3\,s}.$

\subsection{Characterizing the Turbulence}

When studying low velocity $(|\mathbf{v}_{T}|\ll$ velocity of sound) turbulence, 
it is usually assumed that the fluid is incompressible 
$(\nabla\mathbf{\cdot}\mathbf{v}_{T}=0).$
The functions $T_{NN}$ and $T_{LL}$ are, then, related in the way described
by Subramanian \citep{subramanian99}. When the above approximation is not valid
(as is the case here), $\mathbf{\nabla\times {v}}_{T}=0$ is used and these functions are
related by $T_{LL}=T_{NN}+r\,(d\,T_{NN}/d\,r)$ \citep{monin75}. The fluid flow 
correlation functions can be written as 
\begin{eqnarray}
2C\left( r\right) &=&\frac{\Omega L_{c}^{2}}{h}\left[ 1-\left( \frac{r}{L_{c}
} \right) ^{q}\,\right] \qquad 0<r<L_{c}  \label{j}, \\
T_{NN}\left( r\right) &=&A_{N}\left[ 1-\left( \frac{r}{L_{c}}\right) ^{p}\, 
\right] \qquad l_{c}<r<L_{c}  \label{k}, \\
T_{NN}\left( r\right) &=& 0\qquad r>L_{c},  \label{l}
\end{eqnarray}
with $A_N=V_{c}\,L_{c}\,/3$ \citep{vainshtein82}.
In our study, $l_c$ is much smaller than the numerical
resolution used. We, therefore, considered $M_L\left( 0\right) = M_L\left(
l_c\right).$   For free turbulence, we have $p=4/3,$ (the Richardson law) 
\citep{macomb}. We take a range of values for $p$
to take into account the uncertainties in the physics of high redshift objects
$1\leq p \leq 2.$ It is customary to take $q=2$ for the helicity spectrum,
but here, we shall allow for a more general dependence, using $q \geq 1.$
\par
The required integration time can be estimated from the fact that when the
kinetic energy density of the turbulence equals the magnetic energy density
$(\rm{i.e.,} \;\rho_n\,V_{c}\,/2\sim M_L\left( l_c\right) ),$ turbulence cannot supply
more energy to create stronger magnetic fields. In the integrations that we
performed, the growth saturated before this condition was reached.

\subsection{Discussion}

Both the size of the coherent region $L_M$ and the induced intensity of the magnetic field
$B_M$ were studied. We estimated the value of $B_M$ at all points where $M_L > 0$ 
as $B_M\left( r\right) \sim M_L\left( r\right) /M_L^{\,1/2}\left(0\right).$
We found that, in general, the magnetic correlations that result from the
evolution of the turbulent kinematical dynamo in going from $z=10$ to $z=5,$ are 
independent of the initial field correlations.
\par
We investigated the following sets of turbulent parameters: 1) $L_c=33\,\rm{pc}$ and
$V_c=45\,\rm{km\,s^{-1}};$ and  2) $L_c=81\;\rm{pc}$ and
$V_c=96\,\rm{km\,s^{-1}}.$  In
both cases, we took $p=4/3,$ $q=2,$ $\eta \sim 5\times 10^{\,3}\,\rm{cm^{\,2}\,s^{-1}},$
and $a \simeq 9.76\times 10^{\,39}\,\rm{cm^{\,3}\,s\,g^{-1}}$. For the first set of 
parameters,
we obtained $B_M\sim 1.1\times 10^{-6}\,\rm{G}$ and $L_M\simeq 1.7\,\rm{kpc}$ and for 
the second case, 
$B_M\sim 1.4\times 10^{-6}\,\rm{G}$ and $L_M\sim 5.4\,\rm{kpc}.$
\par
In general, $B_M$ at $z=5$ is sensitive to
the values of $a$ and $V_c.$ The parameter $a$ depends on both $\rho_i$
and $\nu_{in},$ such that it is not possible to discriminate the dependence of our 
results 
on each of these factors, independently. The length $L_M$ depends mainly on the
value of $V_c.$ Different values of $p$ and $q$ change $B_M$ and $L_M$ only slightly.
\par
In Figures (1)-(4), we show $M_L\left(r\right)$ at $z=5$ for 
different values of the parameters. 
\par
In Figure 1, we plotted $M_L\left( r\right)$ as a function of $r$ for 
$p=1.1,$ $1.33,$ and $1.66.$ Using $L_c = 81\,\rm{pc},$ $V_c = 96\, \rm{km\,s^{-1}},$
$a = 9.76\times 10^{\,39}\,\rm{cm^{\,3}\,s\,g^{-1}},$ 
and $q=2,$ we see that $B_M$ and $L_M$ are somewhat larger for smaller values of $p.$ 
\begin{figure}
\resizebox{\hsize}{!}{\includegraphics{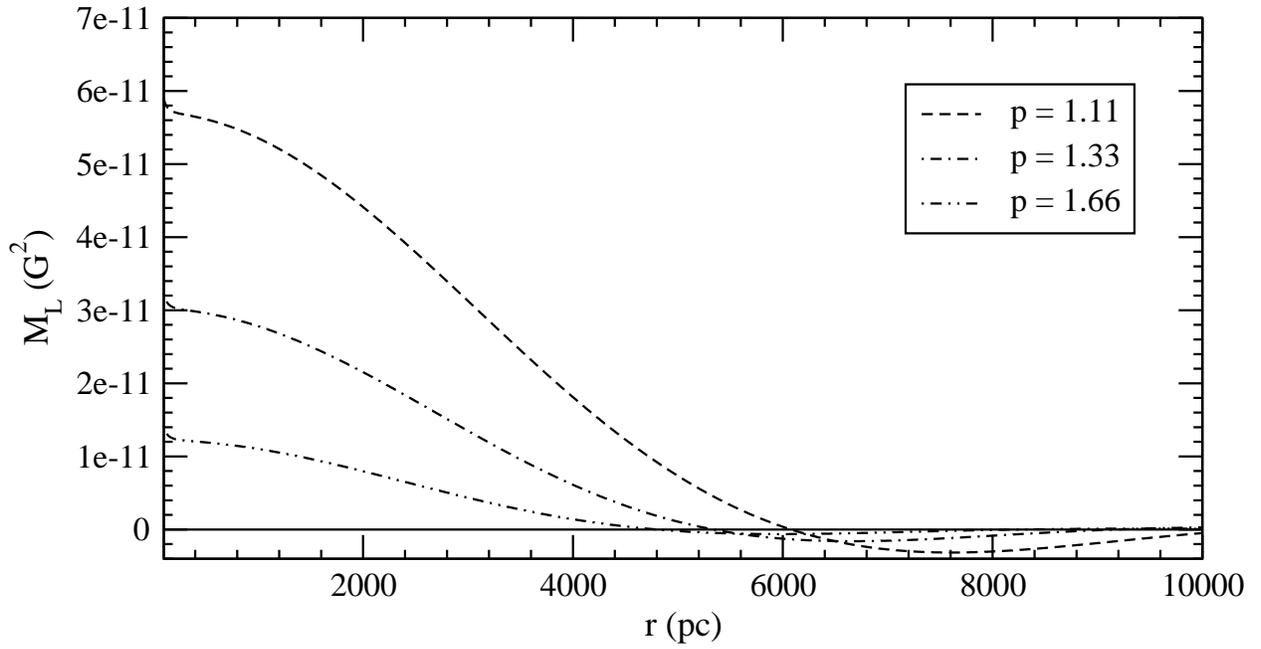}}
\caption{Final value of $M_L(\rm{G^{\,2}})$ as a function of $r(\rm{pc})$ 
for $p=1.11,1.33,$ and $1.66;$ $q=2;$ $L_c = 81\,\rm{pc};$ $V_c=98\,\rm{km\,s^{-1}};$ 
and 
$a = 9.76\times 10^{\,38}\,\rm{cm^{\,3}\,s\,g^{-1}}.$ We see that $B_M$ and $L_M$ 
are somewhat larger for smaller values of $p.$}
\end{figure}
\par
In Figure 2, we show $M_L\left( r\right)$ as a function of $r$ for 
$q=1.8,$ $2,$ and $2.5,$ using $p=1.33$ and the same values for $L_c,$
$V_c,$ and $a$ as in Figure 1. The values of $B_M$ and $L_M$ are a little 
larger for smaller values of $q.$ 
\begin{figure}
\resizebox{\hsize}{!}{\includegraphics{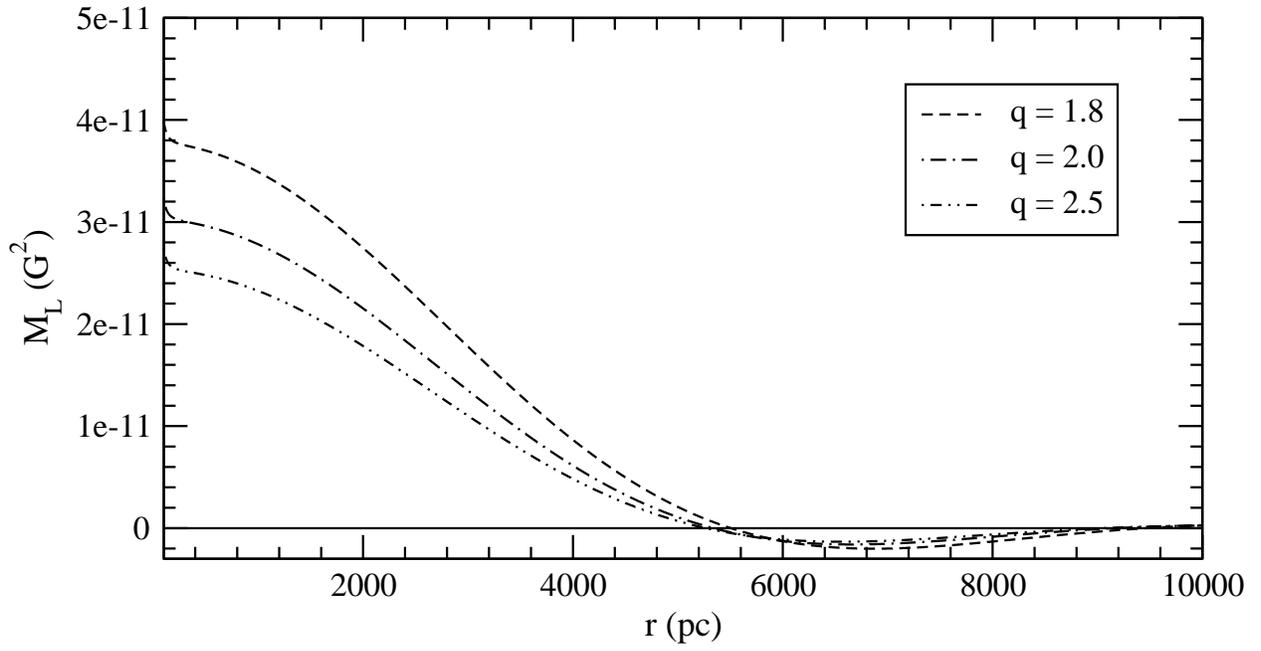}}
\caption{Final $M_L(\rm{G^{\,2}})$ as a function of $r(\rm{pc})$ for 
$q=1.5,2,$ and $2.5;$ $p=1.333;$ $L_c = 81\,\rm{pc};$ 
$V_c=98\,\rm{{km}\,s};$ $a = 9.76\times 10^{\,38}\,\rm{cm^{\,3}\,s\,g^{-1}};$
and ${p=1.33}.$ For smaller values of $q,$ $B_M$ and $L_M$ are somewhat larger. }
\end{figure}
\par
In Figure 3, we plotted $M_L\left( r\right)$ as a
function of $r$ for $a = 3.2\times 10^{\,39}\;\rm{ cm^{\,3}\, s \,g^{-1}},9.7\times
10^{\,39}\;\rm{cm^{\,3}\, s\, g^{-1}},$ and $2.43 \times 10^{\,40}\;\rm{cm^{\,3}}\, 
s \,g^{-1};$
$p=1.33;$ and $q=2.$ We used the same values for $L_c$ and $V_c$ as in Fig. 1. 
We see that the smaller the
value of $a$ (high ion density and/or high ion-neutral collision
frequency), the larger the value of $B_M$. 
However, $L_M$ is almost insensitive to the value of $a$. 
\begin{figure}
\resizebox{\hsize}{!}{\includegraphics{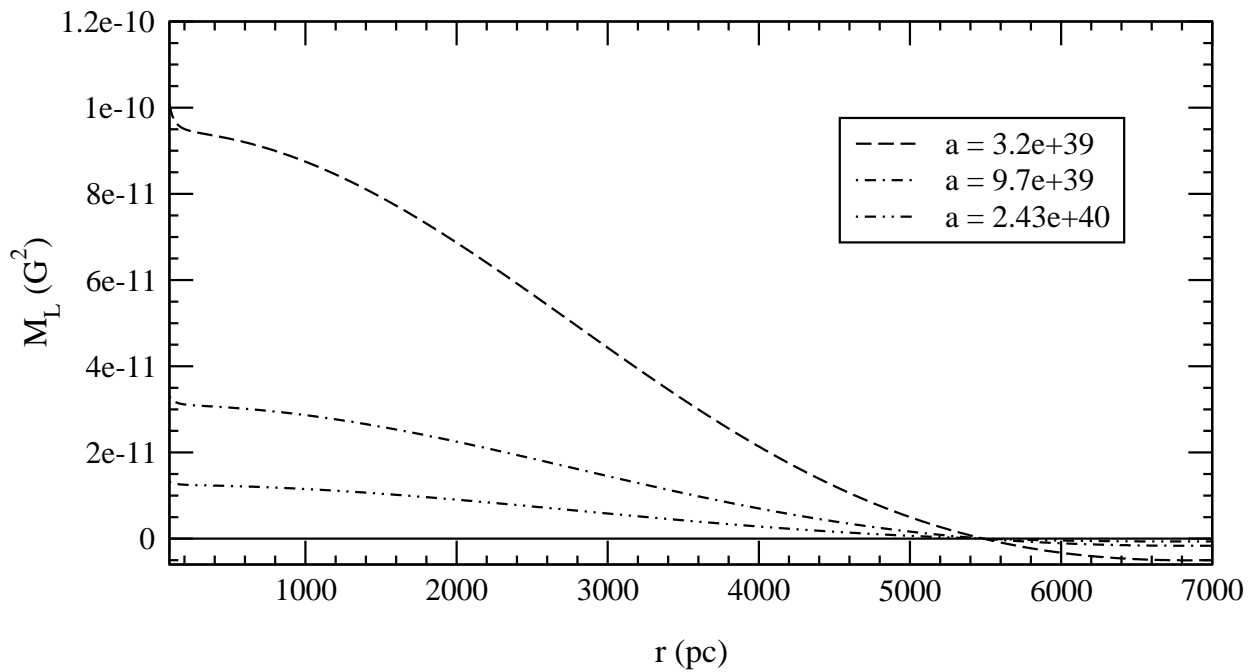}}
\caption{Final values of 
$M_L\rm({G^{\,2})}$ as a function of $r(\rm{pc})$   for 
$q=2,$ $p=1.333,$ $L_c = 81\,\rm{pc},$ 
and $V_c=98\,\rm{{km}\,s^{-1}}$ and three different values for $a:$
$3.2\times 10^{\,38}\,\rm{cm^{\,3}}\, s\, g^{-1},$ 
$9.76\times 10^{\,38}\,\rm{cm^{\,3}\, s\, g^{-1}},$ and 
$2.43 \times 10^{\,39}\,\rm{cm^{\,3}\,s\, g^{-1}}.$ For
smaller values of $a$, $B_M$ is larger,
while $L_M$ remains practically unchanged.} 
\end{figure}
\par
In Figure 4, we plotted $M_L\left( r\right)$ as a
function of $r$ for $q=2,$ $p=1.33,$ $L_c = 81\,\rm{pc},$ $a = 3.2\times 10^{\,39}\,\rm{
cm^{\,3}\,s\, g^{-1}},$  and $V_c = 30\,\rm{ km \,s^{-1}},$ $45\,\rm{km\,s^{-1}},$ and $96
\,\rm{{km}\, s^{-1}}.$
We see that large values of these two parameters produce high values of $M_L$
as well as large coherence lengths.
\begin{figure}
\resizebox{\hsize}{!}{\includegraphics{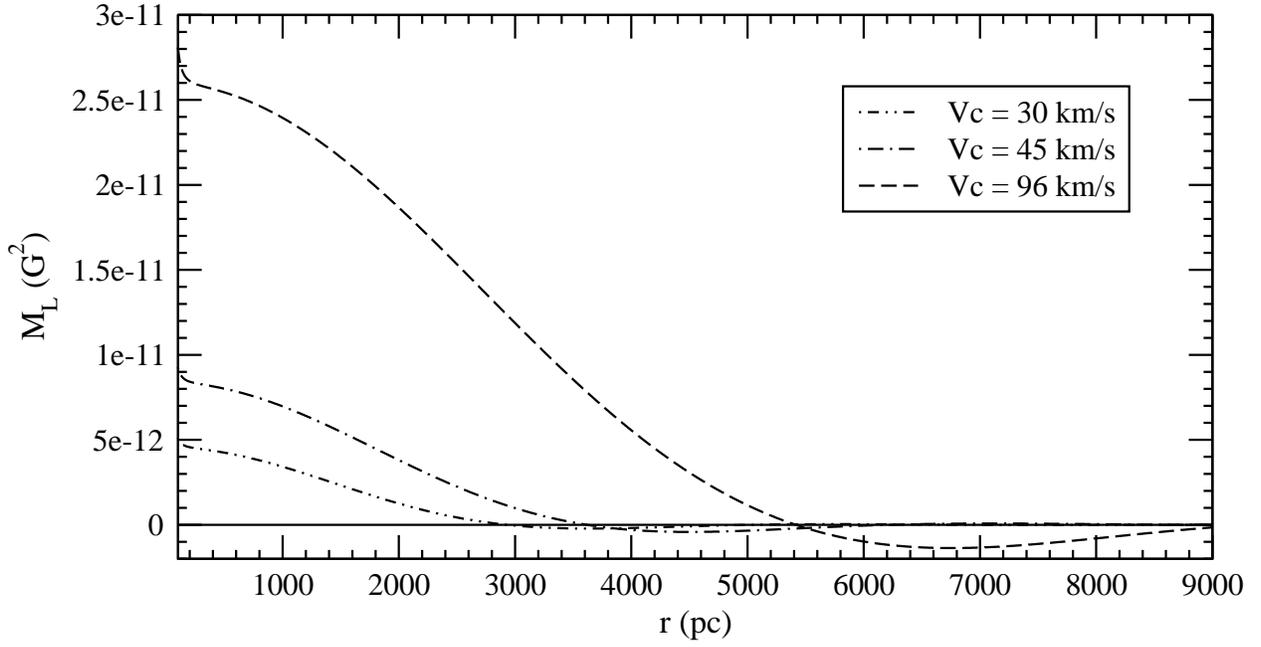}}
\caption{Final values of 
$M_L\rm{(G^{\,2}})$ as a function of $r(\rm{pc})$ 
for $q=2,$ $p=1.33,$ $L_c = 81\,\rm{pc},$ and $a = 9.76\times 10^{\,38}\,\rm{cm^{\,3}\,
s\, g^{-1}}$ 
and three different values for $V_c$:$ 30\,\rm{km\,s^{-1}}$, $45\,\rm{ km\, s^{-1}}$,
and $96\,\rm{{km}\,s^{-1}}.$ 
We see that for larger values of $V_c$, $B_M$ and $L_M$ are larger.}
\end{figure}
\par
Finally, in Figure 5, we plotted the evolution of $M_L$ as a function of $t$
at $r_0 = 112\;\rm{pc},$ going from $z=10$ to $z=5,$
for $M_L\left( r_0,\,t=0\right) = 1.5\times 10^{-38}\;\rm{ G
^{\,2}}$, $1.5\times 10^{-47}\;\rm{ G^{\,2}},$ and $1.5\times 10^{-55}\;\rm{ G^{\,2}}$. 
We see that after a short period of time, $t\sim 5\times 10^{\,6}\;\rm{ years},
$ $M_L$ reaches its saturation value, $M_L^{\,1/2}\sim 10^{-6}\;\rm{ G},$ 
independent of its initial value. 
\begin{figure}
\resizebox{\hsize}{!}{\includegraphics{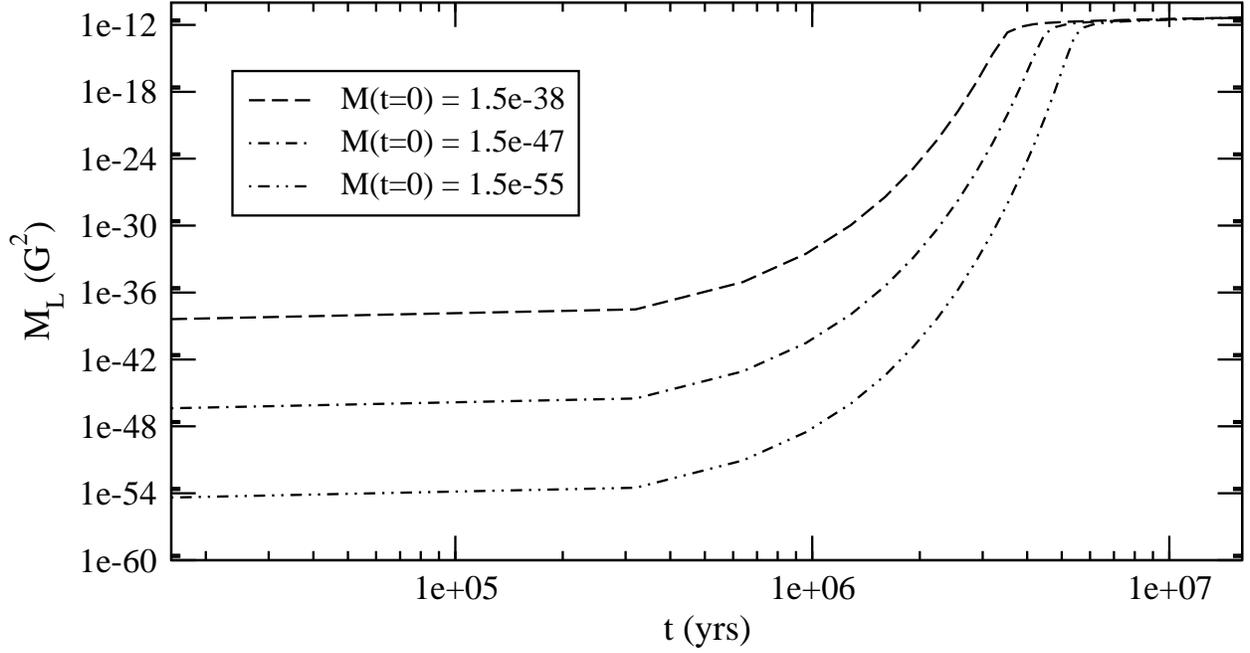}}
\caption{Values of $M_L\rm{(G^{\,2})}$
as a function of $t\rm{(years)}$  at $r_0=112\;\rm{pc}$ for three 
different initial values: $M_L\left( r_0,t=0\right) =
1.5\times 10^{-38}\,\rm{ G^{\,2}},$ 
$1.5\times 10^{-47}\,\rm{ G^{\,2}},$ and $1.5\times 10^{-55}\,\rm{ G^{\,2}}.$ We used 
$p=1.11,$ $q=2,$ $L_c = 81\,\rm{pc},$
$V_c=98\,\rm{km\,s},$ and
$a = 9.76\times 10^{\,38}\,\rm{ cm^{\,3}\, s \,g^{-1}}.$ Independent of the
initial values of $M_L,$ saturation occurs at a time 
$t\sim 5\times 10^{\,6}\,\rm{ years},$
when $M_L^{\,1/2}\sim 10^{-6}\,\rm{ G}$.}
\end{figure}
\par
Throughout the integration time, the kinetic energy density of the fluid was greater 
than that of the magnetic energy. The magnetic energy density is given by $E_B = 
B^{\,2}/8\,\pi ,$ which for $B \sim
10^{-6}\;\rm{G}$ (see Fig. 5), has a value of  $E_B \simeq 10^{-18}\;\rm{ erg\,cm^{-3}}.$ 
The kinetic energy density is given by $E_V = \rho_nV_c^{\,2/}2.$  For
$V_c = 98 \;\rm{km \,s^{-1}}$  and the range of values used for $\rho_n,$ the kinetic
energy, $3\times 10^{-13}\;\rm {erg\,cm^{-3}}$ 
$\la E_V \la 10^{-9}\;\rm{ erg \,cm^{-3}},$ was many orders of magnitude greater than 
the magnetic energy. 

\section{CONCLUSIONS}

In this work, we studied the problem of the origin of strong coherent large-scale magnetic 
fields, observed in low-redshift galaxies and previously thought to have been created by 
the mean field dynamo. Since doubts have been cast on the mean field dynamo as the source 
of these fields due to their observation in high redshift objects, where the dynamo would 
not have had suficient time to operate, we investigated here the stochastic helical dynamo 
as such a source. We showed that these fields  can be generated in the plasmas found in 
very high
redshift objects.
\par
The generation of strong large scale coherent fields is a highly non-linear
magnetohydrodynamical problem, which depends upon many factors. 
Here we discussed
a possible mechanism for the generation of large scale fields, namely the non-linear 
evolution and diffusion of magnetic noise. Magnetic noise  becomes coherent on a
scale which is larger than that of the turbulence due to the presence of non
linear terms in the evolution equations for the magnetic correlations.
\par
We found that for realistic turbulent parameters, it
is possible to generate the magnetic fields, observed at high redshifts .
Between $z\sim 10$ and $z\sim 5$, the primordial plasma is strongly turbulent and 
partially ionized
The magnetic field intensities reached are independent of initial correlations. 
\par
We considered a very simple model for the
generation and evolution of the magnetic correlations. The dependence of
the resulting magnetic field intensity on the charge composition of the
primordial plasma, suggest that the reionization and star formation processes 
played an important role in determining the features of the
magnetic fields detected in high redshift objects. In a forthcoming work we
shall address the evolution of the magnetic correlations, considering a time
dependent ion density as well as other nonlinear processes, such as the the
Hall effect. Other turbulent scenarios, in addition to the
homogeneous and isotropic one considered here, will be treated as well.

\section{ACKNOWLEGEMENTS}

We thank George Morales for useful comments. This work was partially supported 
by the Brazilian financing agency FAPESP
(00/06770-2). A.K. acknowledges the FAPESP fellowship (01/07748-3). R. O.
acknowledges partial support from the Brazilian financing agency CNPq
(300414/82-0).

\end{document}